\documentstyle[aps,prl]{revtex}
\begin{document}
\draft
\twocolumn[\hsize\textwidth\columnwidth\hsize\csname %
@twocolumnfalse\endcsname

\title{ Thermodynamic properties of the planar $t-J$ model}
\author{ J. Jakli\v c and P. Prelov\v sek }
\address{ J. Stefan Institute, University of Ljubljana, 
1001 Ljubljana, Slovenia }
\date{\today}
\maketitle
\begin{abstract}\widetext
Several thermodynamic quantities within the planar $t-J$ model are
calculated using the $T>0$ Lanczos method on clusters of up to 26
sites. Hole density $c_h(\mu,T)$ shows a non-Fermi liquid behavior as
a function of $T$ and suggests a transition from small to large Fermi
surface at $c_h\sim 0.15$. Specific heat reveals a maximum at the
exchange energy-scale up to $c_h=0.2$, where it becomes almost
$T$-independent for $T\agt 0.15~t$.  At constant $T$ the entropy has
maximum for $c_h\sim 0.15$, with large values at low $T$, consistent
with experiments on cuprates. In the underdoped regime the spin
susceptibility $\chi_0(T)$ exhibits a maximum at finite $T = T^*$,
with $T^*$ decreasing with doping, and disappearing for $c_h>0.15$.

\end{abstract}
] 
\narrowtext 
The normal state of the superconducting cuprates, in particular the
character of the low-energy excitations, still lacks proper
understanding. Although there appears to be a consolidation of
experimental results probing their transport \cite{batlogg}, magnetic
\cite{torance}, thermodynamic \cite{loram}, and single-electron 
\cite{shen} properties, on the theoretical side several important 
questions about the nature of the corresponding quantum liquid remain
open.  While undoped cuprates are well understood in terms of a
two-dimensional (2D) $S=1/2$ antiferromagnet (AFM), it is less clear
what changes are induced on doping the AFM with mobile holes. Besides
the questionable existence of quasiparticles (QP) in the sense of some
type of Fermi liquid, the character of the QP and the associated Fermi
surface (FS) in doped AFM are also controversial. While at low doping
theoretical studies \cite{trugman,eder} as well as experimental
findings \cite{wells} seem to favor the interpretation in terms of
hole QP in a rigid band and related hole-pocket FS, situation at
larger doping is more consistent with the usual large electronic FS
\cite{stephan,singh2,shen}. Nevertheless, several
normal-state dynamical properties of cuprates have recently been 
reproduced by using unbiased numerical techniques with strong
correlations taken fully into account \cite{jplanc,jpdyna}. It is the
aim of this paper to present results on relevant thermodynamic
properties of correlated electrons, which could help in finding a
proper concept for such systems, but also in establishing the relevant
model for cuprates.

In the following we address some of the questions above by studying
finite-$T$ static properties of the 2D $t-J$ model, as a prototype
model for strongly correlated electrons and cuprates in particular
\cite{rice}
\begin{equation}
H=-t\sum_{\langle ij\rangle  s}(c^\dagger_{js}c_{is}+ \text{H.c.})
+J\sum_{\langle ij\rangle} (\vec S_i\cdot \vec S_j - 
{1\over 4} n_i n_j)~. \label{model}
\end{equation}
Here $c^\dagger_{is}$, $c_{is}$ are projected fermionic operators,
excluding the states with doubly-occupied sites. To investigate the
regime of cuprates we set $J/t=0.3$.

Although there have been so far extensive numerical studies of the
ground-state properties of the $t-J$ model \cite{dagorev}, its
finite-$T$ properties are less explored. AT $T>0$ a fruitful technique
in addressing several thermodynamic properties proved to be the
high-$T$ series expansion \cite{singh,putikka,singh2}. In the present
study we aim at the calculation of the finite-$T$ static properties of
the model Eq.~(\ref{model}) with a recently introduced method,
combining Lanczos diagonalization technique and random sampling
\cite{jplanc}, which has been so far applied to the study of several
dynamical responses \cite{jpdyna}. We concentrate on calculating the
quantities, which are related to the expectation values of the
conserved operators, commuting with the Hamiltonian, Eq.(1). Such
operators are e.g. total number of electrons $N_e$, total energy $H$,
and $S_z$, the $z$-component of the total spin operator. In terms of
them some interesting thermodynamic quantities can be expressed, in
particular chemical potential $\mu$, entropy $S$, specific heat $C_V$
and the uniform susceptibility $\chi_0$.  The reason to limit our
treatment only to commuting operators is technical, since in this case
we can avoid time and memory consuming computation of wavefunctions
and their scalar products. This in turn enables us to study larger
clusters with up to 20 sites at all doping concentrations.

We follow the method for the evaluation of $T>0$ static quantities
\cite{jplanc}. To calculate the quantity of the type $\text{\rm{Tr}}
[f(N_e,S_z,H)e^{-\beta(H-\mu N_e)}]$, the trace is expanded in terms
of an arbitrary orthonormal set of basis functions $|n\rangle$,
$n=1\ldots K$, with good quantum numbers $N_e$ and $S_z$. Each state
$|n\rangle$ is used as initial function $|\phi_0^n\rangle$ for the
$M$-step Lanczos procedure \cite{dagorev}, yielding a subspace spanned
by functions $|\phi_j^n\rangle$, $j=0\ldots M$. The Hamiltonian is
diagonalized in this subspace to obtain the eigenvalues $E_j^n$ and
corresponding eigenvectors $|\psi_j^n\rangle$, which are used to
evaluate $\langle f \rangle$
\begin{equation}
\langle f\rangle \approx \frac{N_{st}}{K\Omega }\sum_{n=1}^{K}
\sum_{j=0}^{M-1}|\langle n|\psi_j^n \rangle|^2 f(N_e^n,S_z^n,E_j^n) 
e^{-\beta(E_j^n-\mu N_e^n)},
\label{method}
\end{equation}
where $\Omega=\text{\rm{Tr}}[e^{-\beta(H-\mu N_e)}]$ is the
grand-canonical partition function. Clearly, the expression
Eq.(\ref{method}) is exact for $K=M=N_{st}$, where $N_{st}$ is the
dimension of the complete basis. In practice $K,M\ll N_{st}$ are used,
with states $|n\rangle$ sampled randomly. More detailed discussion of
the method and tests are given elsewhere \cite{jplanc,jpdyna}. In the
present application it is essential that scalar products $\langle
n|\psi_j^n\rangle$ are trivially evaluated, since this is the
component of the function $|\phi_0^n\rangle$ in the wavefunction
$|\psi_j^n \rangle$, i.e. can be obtained from the eigenvectors of the
$M\times M$ tridiagonal matrix without storing $|\phi_j^n\rangle$.
Typically also small $M\alt 100$ is used, thus reorthogonalization of
the Lanczos functions can be avoided.  The computational effort in the
present case is then equal to a ground-state Lanczos procedure,
repeated $K$ times. We employ in the following typically $K\sim
200-1000$ in every $N_e$ sector.

We have calculated a few static quantities above using the
approximation Eq.~(\ref{method}) on clusters with $N=16$, $18$, and
$20$ sites, thus having control over the finite-size effects through
comparison of the results. We also treated the undoped system with
$N=26$ sites. In general the results are quite free of finite-size
effects for temperatures $T>T_{\text{fs}}(N,c_h)$, which amounts to
$T_{\text{fs}}\sim 0.1~t$ for $N=20$ and $c_h\le 0.3$.
 
We first analyse the hole density $c_h=1-\langle N_e\rangle/N$ as a
function of $T$ and $\mu$. In Fig.~\ref{fig1}(a) we present curves
$c_h(T)$ for several $\mu_h$, where the chemical potential for holes
is $\mu_h=-\mu$. Clearly, for $\mu_h<\mu_h^0$ the ground state of the
system contains no holes. Here $\mu_h^0= -1.99~t$ \cite{dagorev} is
related to the minimum energy of a single hole added into the undoped
system. The corresponding curves for $\mu_h<\mu_h^0$ are denoted with
dashed lines.  In Fig.~\ref{fig1}(a) we in general do not find a $T^2$
dependence of $c_h$ at low $T$, as expected for a normal Fermi liquid,
except within the extreme overdoped regime $c_h>0.3$. In particular,
in a broad range $0.15 <c_h <0.3$ very unusual linear variation of
$c_h(T)$ is observed at lowest $T$. Moreover, a remarkable feature is
a non-monotonous $c_h(T)$ dependence as the value $c_h(0)$ is
varied. There exists a doping concentration $c_h^*$ such that
derivative $dc_h/dT$ at low $T$ is positive for curves with $c_h(T\sim
0)< c_h^*$ and negative for curves with $c_h(T\sim 0)> c_h^*$. The
marginal doping $c_h^*\sim 0.15$ seems to be system
independent, as checked quantitatively for systems with $N=16$, $18$,
$20$.

In Fig.~\ref{fig1}(b) we plot the variation of $c_h$ with $\mu_h$ at
several chosen temperatures $T$. We include also 
$T=0.05~t<T_{\text{fs}}$,
since in this case it shows less pronounced finite-size effects. As
$T\to 0$ clearly $c_h(\mu_h<\mu_h^0)\to 0$, while
$c_h(\mu_h>\mu_h^0)$ remains finite. The slope $dc_h/d\mu_h$, being
proportional to the compressibility $\kappa$ of the hole fluid, is
finite for $T>0$, indicating the absence of the phase separation in
the system at chosen $J/t$ \cite{putikka}. On the other hand,
Fig.~\ref{fig1}(b) reveals that for $T\to 0$ $\kappa$ is increasing
and becoming large on approaching $\mu_h \sim \mu_h^0$, consistent
with large (polaron) mass enhancement of individual holes at vanishing
doping. Moreover we cannot exclude a singular behavior (divergence)
of $\kappa$, as deduced from analogous results for the Hubbard model
\cite{imada}.

Next we find remarkable a feature, appearing at the hole-density of
$c_h\sim c_h^*$, where $\mu_h$ is essentially pinned at the value
$\mu_h\sim -1.8~t$, i.e. does not vary with $T$. This pinning is
active in a wide range of $T$, and is again almost not dependent on
the system size.  It is tempting to interpret the marginal
concentration $c_h^*$ as a change of the character of the FS. To
establish the relation, we still have to rely on arguments which apply
to the gas of non-interacting fermions. Denoting by $g(\varepsilon)$
the density of the single-fermion states, a simple Sommerfeld
expansion yields that the number of particles at fixed chemical
potential $\mu$ is given by $c_e(T)=c_e(T=0)+(\pi^2/6)(k_BT)^2
g'(\mu)$.  Therefore, the prefactor of the $T^2$ term is just
proportional to the derivative of the density of states. Indirectly
this gives an information about the character of the FS, since one
would plausibly associate $g'(\mu)>0$ for $c_e\alt 1$ with a large
electron FS, and oppositely $g'(\mu)<0$ with hole-like FS or small
hole pockets vanishing for $c_e \to 1$. For 2D it is straightforward
to represent $g'(\mu)$ more explicitly as a line integral
\begin{equation}
g'(\mu)=\frac{1}{2\pi^2}\int_{L_{\mu}}
\frac{dl_{\vec{k}}}{|\nabla\varepsilon(\vec{k})|^2}
\bigg[\frac{1}{K}-\frac{\hat{K}\cdot{{\text{\rm\bf m}}^{-1}}\hat{K}}
{|\nabla\varepsilon(\vec{k})|}\bigg]\label{curvature}
\end{equation}
over the curve $L_{\mu}=\{\vec{k};\;\varepsilon(\vec{k})=\mu\}$. Here
$1/K$ is the curvature of $L_{\mu}$ (positive for electron-like orbits
and vice versa), $\hat{K}$ the unit vector perpendicular to $L_{\mu}$,
and ${\text{\rm\bf m}^{-1}}=\partial^2\varepsilon (\vec{k})/ 
\partial \vec k\partial \vec k$ the inverse effective mass tensor. 
Apart from the effective mass term, this is similar to the expression
for the Hall resistivity in terms of the curvature of the FS
\cite{tsuji}. From Eq.~(\ref{curvature}) it follows that at least in
the region of the $\vec{k}$-space, where the effective-mass tensor is
positive-definite, $g'(\mu)>0$ implies also that the average curvature
$K^{-1}$ of the FS $L_\varepsilon$ is positive. Although the observed
non-quadratic $T$-dependence in Fig.~\ref{fig1}(a) questions above
arguments, we may still interpret $dc_h/dT<0$ (i.e. $dc_e/dT>0$) in
Fig.~\ref{fig1}(a) as an indication for $g'(\mu)>0$, i.e.  positive
average curvature of the FS for $c_h \agt c_h^*$. This in turn implies
a transition from the hole-pocket picture \cite{trugman,eder,wells},
existing at low doping, to an electron-like FS
\cite{stephan,singh2} at  $c_h \sim c_h^*$.

Next we consider the entropy
\begin{equation}
S=k_B\ln\Omega+(\langle H\rangle-\mu\langle N_e\rangle)/T.
\label{entropy}
\end{equation}
As a test we have checked and found agreement of our entropy data with
available results from the high-$T$ series \cite{putikkac}. In
Fig.~\ref{fig2} we show the doping dependence of the entropy density
(per site) at different $T\le J$ and $N=16-20$. As expected, finite
size effects are most pronounced at lowest $T=0.1~t\sim
T_{\text{fs}}$. Nevertheless, at all $T<J$ the entropy has a broad
maximum at $c_h\sim 0.15$, indicating the highest density of many-body
states in this `optimum' doping regime.

The magnitude of the entropy in the `optimum' regime appears very
large, i.e. at $T=0.1~t=J/3$ the entropy per site is $s\sim 0.39~k_B$,
which is almost 40\% of $s(T=\infty)$ for the same doping $c_h$,
although $T\ll J,t$.  In other words, introducing the degeneracy
temperature by $s(T_{\text{deg}})=s(T=\infty)/2$, for intermediate
dopings of $c_h\sim 0.15$ we get $T_{\text{deg}}\sim 0.17~t$, being
very small in comparison with any resonable QP bandwidth $W$,
e.g. $W=8~t$ of the non-interacting electrons. It is remarkable that
the entropy of such magnitude has been deduced from the electronic
specific heat measurements in $\text{YBa}_2\text{Cu}_3\text{O}_{6+x}$
materials \cite{loram}. E.g., for the optimally doped material with
$x=0.97$ at $T=300~\text{K}$ it was found $\Delta s = 0.35~k_B$ per
planar copper site ($\Delta s = 0.7~k_B$ per unit cell), relative to
the undoped $x=0$ sample. We find the corresponding value
$s(c_h=0.15)-s(c_h=0)\sim 0.30~k_B$ at $T=0.1~t
\sim 450~\text{K}$ (assuming $t=0.4~\text{eV}$ \cite{rice}).

In Fig.~\ref{fig3} we present the $T$-dependence of the specific heat
$C_V=T(\partial S/\partial T)_{\mu}/N$ at different dopings.  For the
undoped AFM we have not observed any appreciable size dependence on
systems with 16 - 26 sites, and our results seem to be even superior to
those obtained by other methods \cite{gomez}. $C_V$ is strongly
$T$-dependent in the observed interval $T=0 - 3.3~J$, with a maximum
at $T\sim 2J/3$, and approximately $C_V \propto T^2$ at low-$T$. As
the system is doped, $C_V(T)$ still exhibits a maximum, which is
however strongly suppressed and gradually moves to lower $T$ with
increasing $c_h$. The peak can be attributed to the thermal activation
of the spin degrees of freedom with an exchange energy scale
persisting in the doped system, as observed already in dynamical spin
correlations \cite{jpdyna}. The presence of the exchange energy scale
however disappears in the `overdoped' regime $c_h\ge 0.3$.

It is characteristic (and consistent with the vanishing role of
$J$) that in the `optimally' doped regime $c_h\sim 0.2$ we find
$C_V(T)\sim \text{const}$ for $0.15~t<T<t$, being far from the Fermi
liquid behavior. The anomalous behavior in this regime implies
$\Omega\propto T$, as follows from Eq.~(\ref{entropy}), as well as a
constant (flat) density of many-body states $D(E) \sim\text{const}$,
as defined through $\Omega=\int D(E) e^{-\beta E}dE$, at least in a
broad intermediate region of energies.  It seems natural to connect
this phenomenon with the observed universality of dynamical response
in the optimum doping regime \cite{jpdyna}.

Only at extreme overdoping $c_h > 0.5$ (low electron density) we were
able to see indications of the Fermi-liquid-like $C_V\propto T$
behavior in a broad interval $T\alt t$.  On the other hand, we have
indications for a different $C_V \sim \gamma T$ behavior at
lowest $T_{\text{fs}}<T<0.15~t$, as deduced from a drop of $C_V$ in
Fig.~\ref{fig3}, which would be closer to the experimental results
\cite{loram} and the QP picture.  Such interpretation would be again
difficult, since it would require $T_{\text{deg}}\sim 800~K$. On the
other hand, it is puzzling that obtained $\gamma$ in the low
$T<0.15~t$ regime can be matched with the value for non-interacting
fermions at $c_e \alt 1$ in a tight-binding band renormalized only by
a modest factor $t/\tilde t \sim 2.5$, as noted previously
\cite{loram}.

As last we consider the uniform spin susceptibility $\chi_0=\langle
(S_z)^2\rangle/Nk_BT$, which is shown in Fig.~\ref{fig4} at several
$c_h$. Again for $c_h=0$ our results agree with other calculations
\cite{singh} down to $T\sim 0.3~J$. In 
the undoped system $\chi_0$ exhibits a maximum at $T=T^*\sim J$, which
gradually shifts to lower $T$ with doping and finally disappears at
$c_h>0.15$. $\chi_0$ at $T\to 0$ increases with doping, in agreement
with experiments \cite{torance} and some previous results
\cite{singh,dagorev}. The existence of such maximum at $T=T^*$ in
undoped cuprates has been interpreted in terms of a pseudogap
\cite{batlogg}. At doping $c_h=0.15$ we observe a monotonous,
Pauli-like $\chi_0$ for $T<0.2~t$, which could signify an onset of a
QP low-$T$ behavior. On the scale of temperatures of the order of $t$,
however, we don't find the Fermi liquid behavior $\chi_0=\text{const}$
up to $c_h>0.6$.

To conclude, we have calculated several thermodynamic properties of
strongly correlated electrons within the $t-J$ model and found
essentially non-Fermi liquid behavior in a wide range of doping
concentrations, up to $c_h > 0.3$. Non-Fermi liquid temperature
dependence is observed in chemical potential, specific heat, and
uniform spin susceptibility, at least in the region of intermediate
temperatures $0.15~t<T<t$, while at very low $T<0.15~t$ we cannot
exclude a possible QP character of low-energy excitations, as
signified by nearly constant $\chi_0(T)$ and $C_V\propto T$.  We
observe in addition a very non-monotonous $c_h(\mu,T)$ dependence,
which we relate to a transition from a small hole-pocket-like to a
large electron-like FS around doping $c_h^*\sim 0.15$. Entropy was
found to be consistent in magnitude with experiments on cuprates,
again showing maximum at $c_h \sim c_h^*$. Finally, the calculated
$\chi_0(T,c_h)$ is consistent with experiments, confirming the $t-J$
model as a suitable one for the explanation of normal-state
properties of cuprates.

\begin{figure}
\caption{(a) Hole density $c_h$ as a function of $T$ at several
values of hole chemical potential $\mu_h$ in steps of $0.1~t$. Dashed
lines denote $\mu_h<\mu_h^0$. (b) $c_h$ as a function of $\mu_h$ at 
various $T$.}
\label{fig1}
\end{figure}
\begin{figure}
\caption{Entropy density $s=S/N$ vs. hole concentration $c_h$ at
several $T$.}
\label{fig2}
\end{figure}
\begin{figure}
\caption{Specific heat per site vs. $T$ at several 
$c_h$. $c_h=0$ was obtained for $N=26$.}
\label{fig3}
\end{figure}
\begin{figure}
\caption{Uniform susceptibility $\chi_0$ vs. $T$ at several $c_h$ in
steps of 0.05. $c_h=0$ was obtained for $N=26$.}
\label{fig4}
\end{figure}


\begin{references}
\bibitem{batlogg} B. Batlogg {\it et al.}, Physica C {\bf 235 - 240}, 
130 (1994).
\bibitem{torance} D. C. Johnston {\it et al.}, Physica C 
{\bf 153-155}, 
572 (1988); Phys. Rev. Lett. {\bf 62}, 957 (1989);
J. B.  Torrance {\it et al.}, Phys. Rev. B {\bf 40}, 8872 (1989).
\bibitem{loram} J. W. Loram, K. A. Mirza, J. R. Cooper, and  W. Y. 
Liang, Phys. Rev. Lett. {\bf 71}, 1740 (1993); J. W. Loram {\it et
al.}, Physica C {\bf 235-240}, 134 (1994).
\bibitem{shen} For a review, see, e.g., Z. X. Shen and D. S. Dessau,
Phys. Rep. {\bf 253}, 1 (1995).
\bibitem{trugman} S. A. Trugman, Phys. Rev. Lett. {\bf 65}, 500 (1990);
E. Dagotto, A. Nazarenko, and M. Boninsegni, {\it ibid.}, {\bf 73},
728 (1994).
\bibitem{eder} R. Eder, Y. Ohta, and T. Shimozato, Phys. Rev. B 
{\bf 50}, 3350 (1994).
\bibitem{wells} B. O. Wells {\it et al}., Phys. Rev. Lett. {\bf 74},
964 (1995).
\bibitem{stephan} W. Stephan and P. Horsch, Phys. Rev. Lett. {\bf 66},
2258 (1991).
\bibitem{singh2} R. R. P. Singh and R. L. Glenister, Phys. Rev. B {\bf
46}, 14313 (1992).
\bibitem{jplanc} J. Jakli\v c and P. Prelov\v sek, Phys. Rev. B 
{\bf 49}, 5056 (1994).
\bibitem{jpdyna} J. Jakli\v c and P. Prelov\v sek, Phys. Rev. Lett. 
{\bf 74}, 3411 (1995); {\it ibid.}, {\bf 75}, 1340 (1995);
Phys. Rev. B {\bf 52}, 6903 (1995);.
\bibitem{rice} For a review, see, e.g., T. M. Rice, in {\it 
Les Houches LVI, 1991}, edited by B. Doucot and J. Zinn-Justin
(Elsevier, 1995), p. 19.
\bibitem{dagorev} For a review, see, e.g., E. Dagotto, Rev. Mod. Phys.
{\bf 66}, 763 (1994).
\bibitem{singh} R. R. P. Singh and R. L. Glenister, Phys. Rev. B {\bf
46}, 11871 (1992).
\bibitem{putikka} W. O. Putikka, M. U. Luchini, and T. M. Rice,
Phys. Rev. Lett.  {\bf 68}, 538 (1992).
\bibitem{imada} N. Furukawa and M. Imada, J. Phys. Soc. Japan 
{\bf 62}, 2557 (1993).
\bibitem{tsuji} M. Tsuji, J. Phys. Soc. Jpn. {\bf 13}, 979 (1958).
\bibitem{putikkac} W. O. Putikka, private communication.
\bibitem{gomez} G. Gomez-Santos, J. D. Joannopoulos, and J. W. Negele,
Phys. Rev. B {\bf 39}, 4435 (1989).


\end{references}
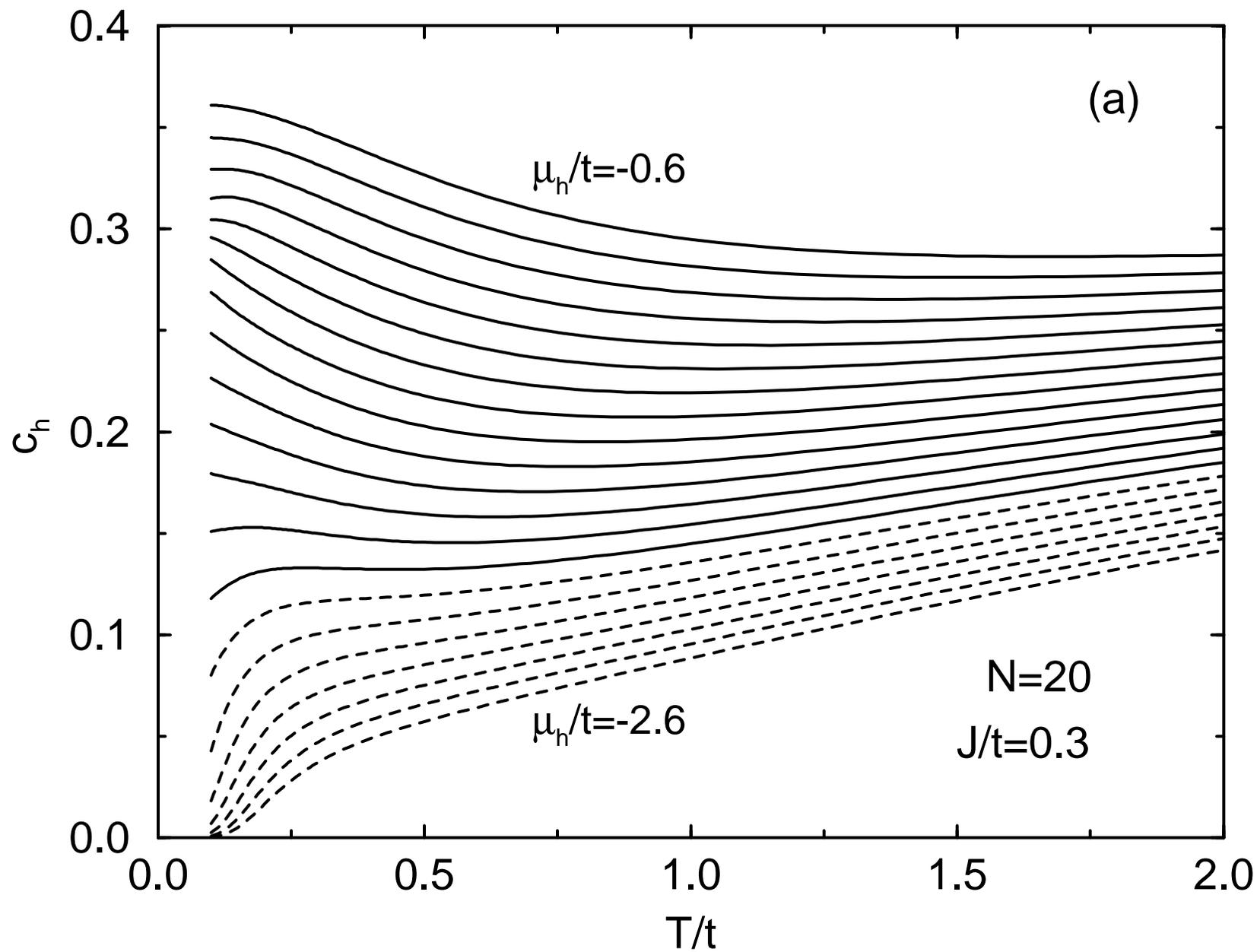
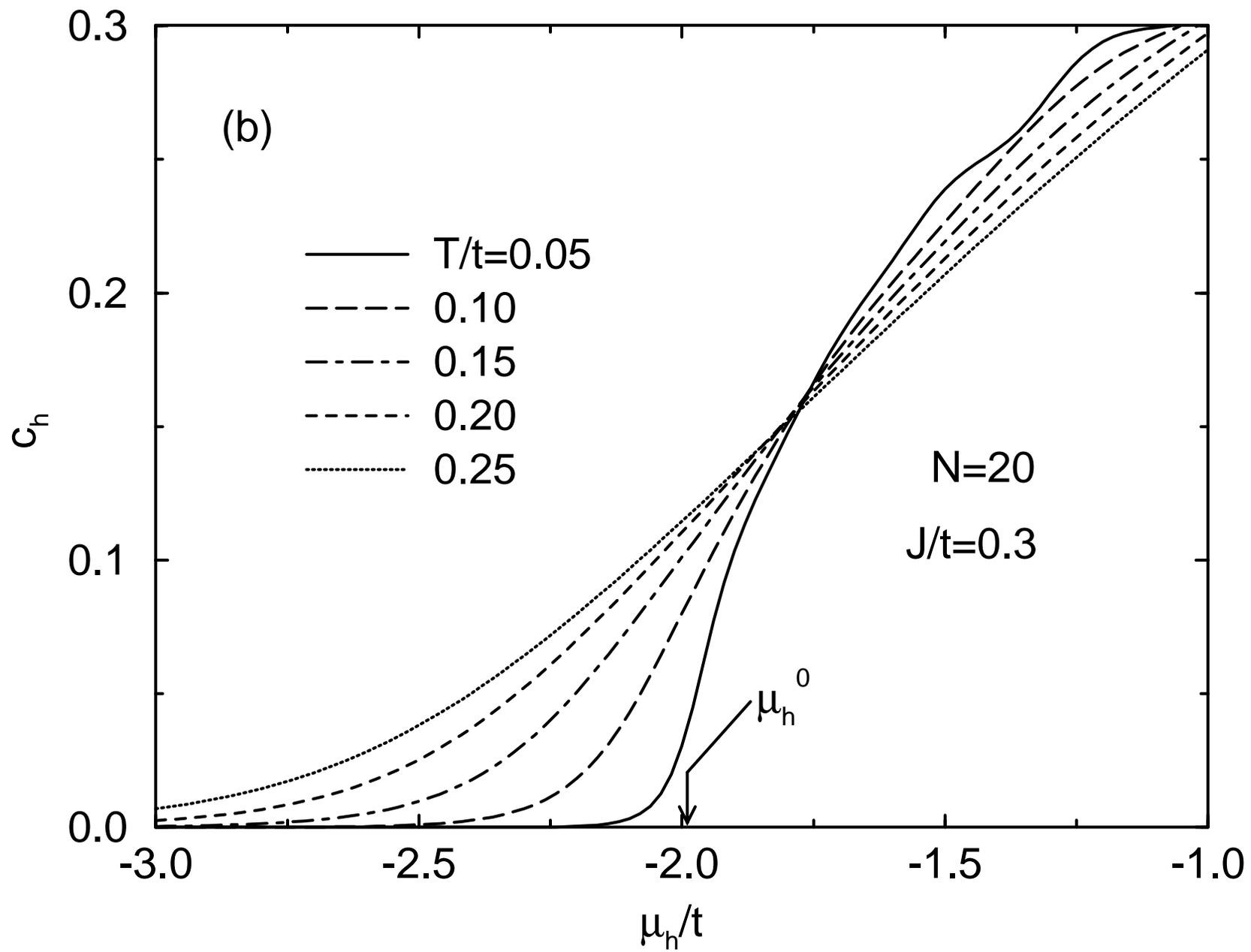
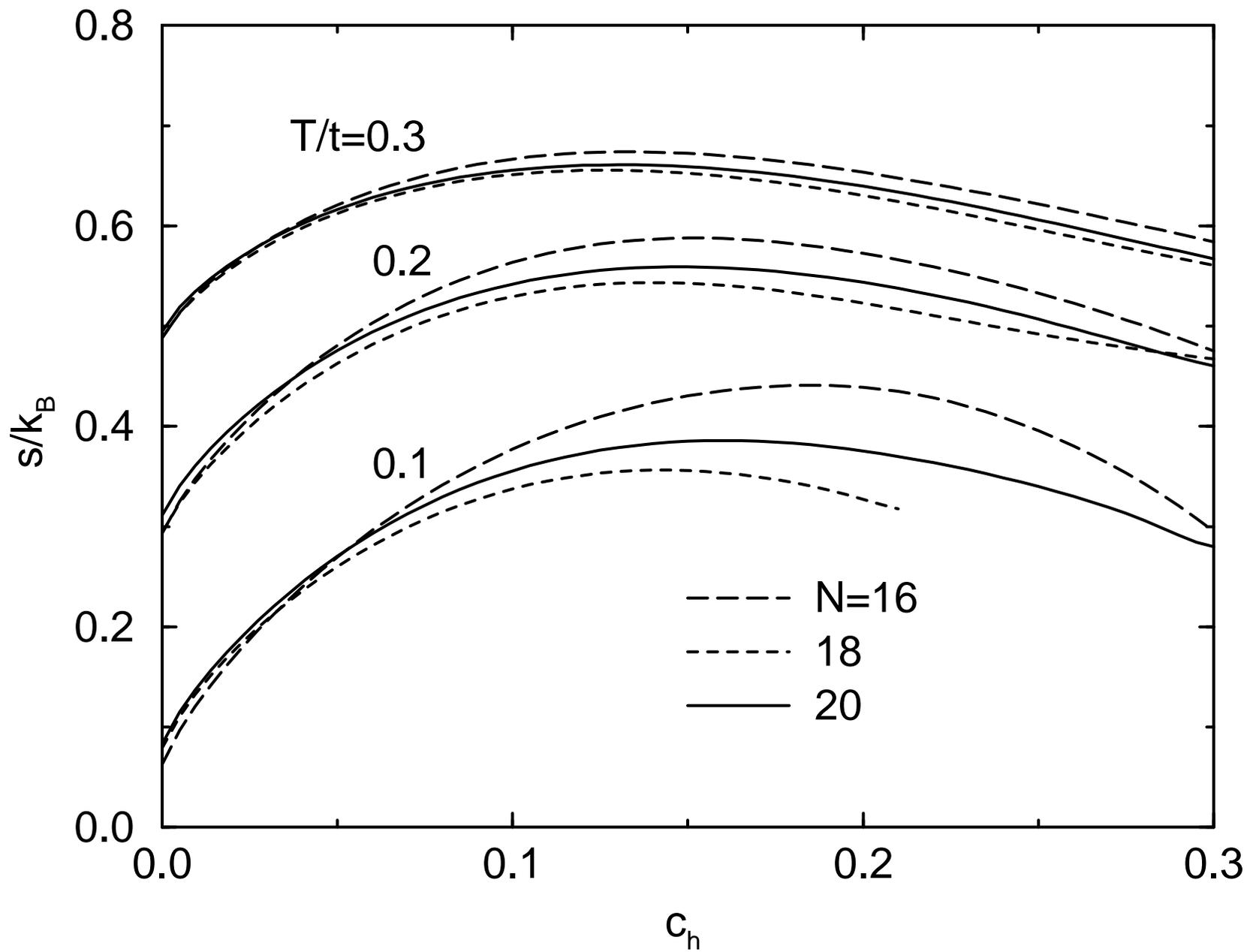
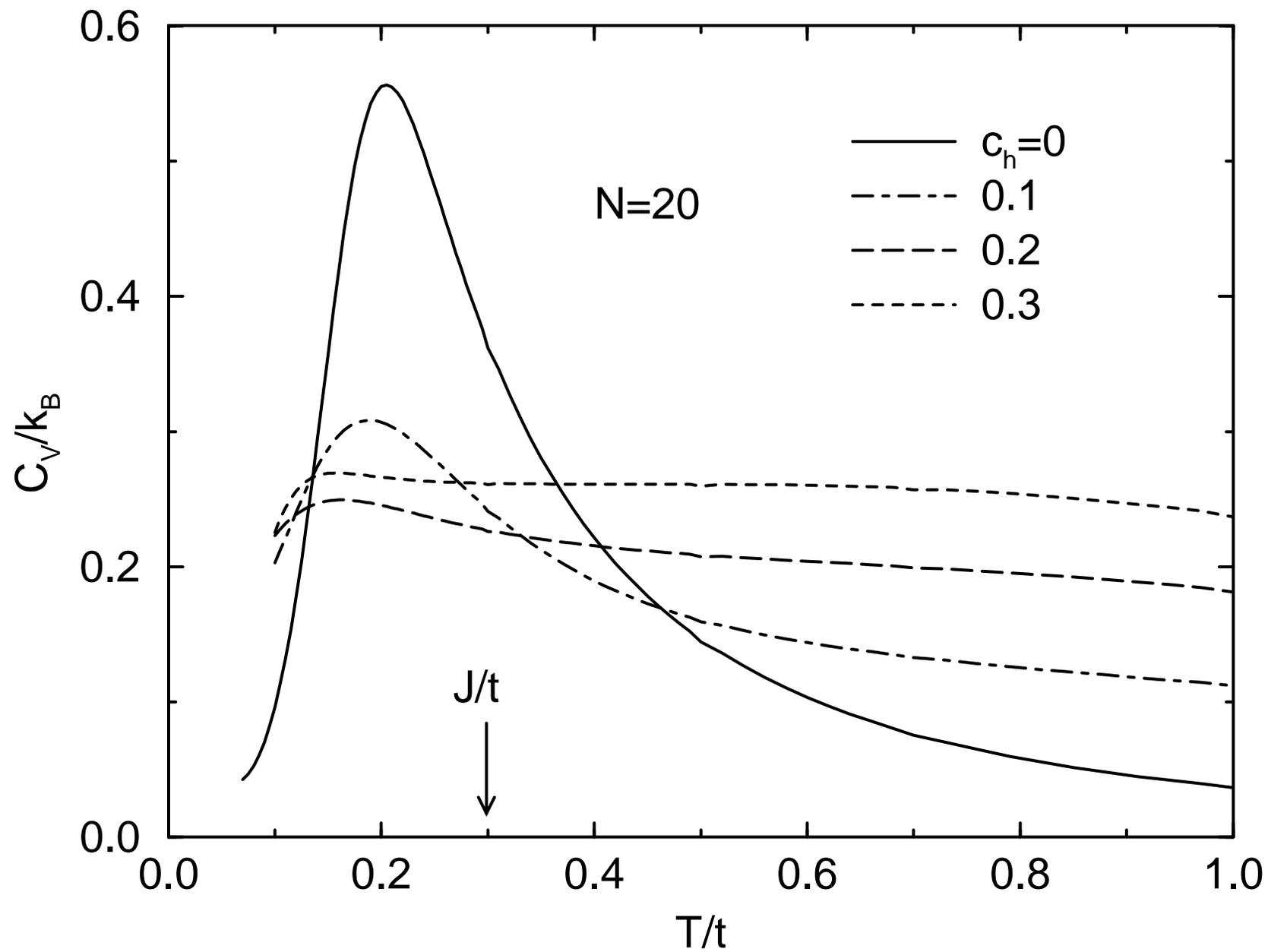
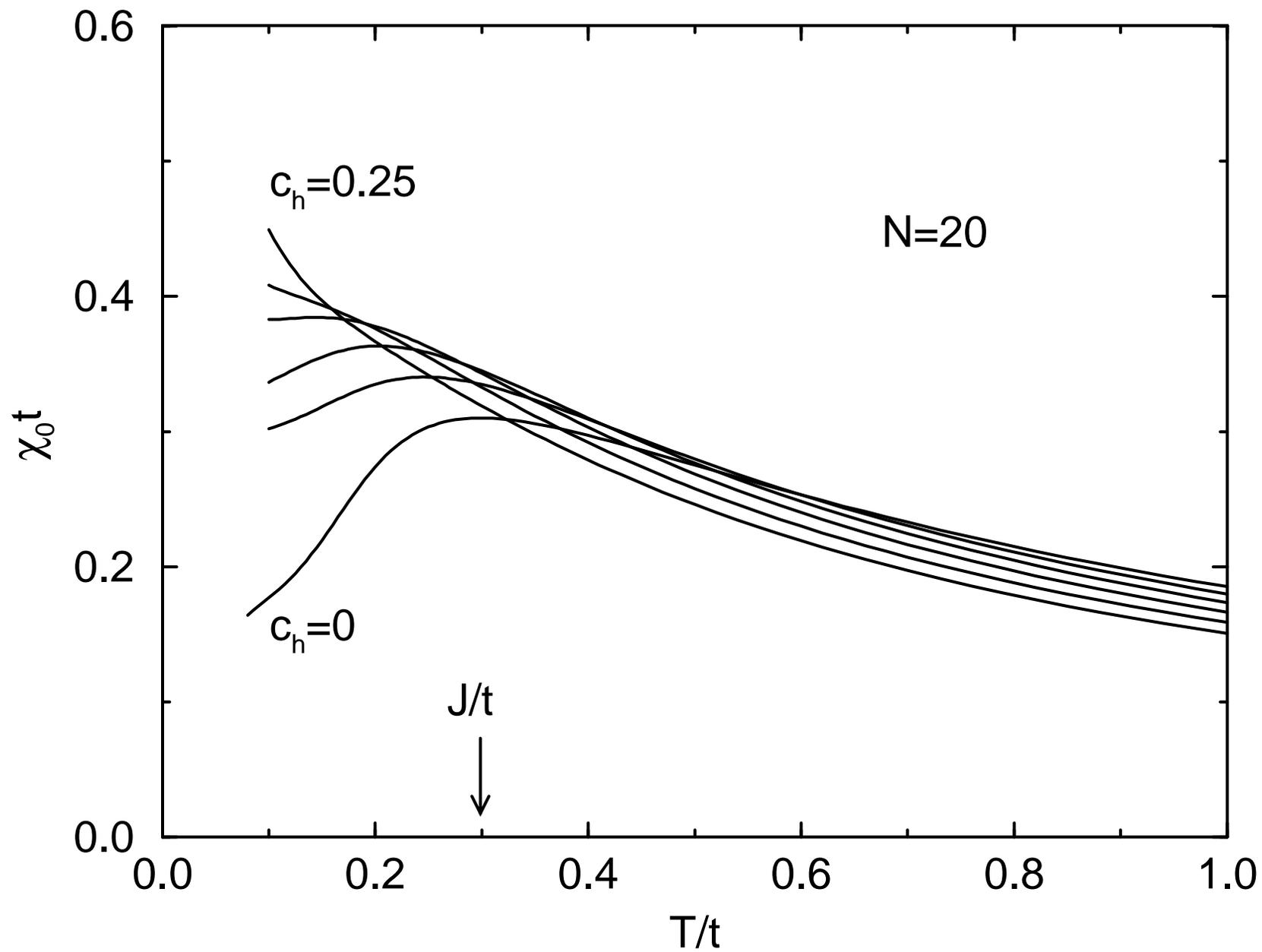
\end{document}